\documentclass[prc,twocolumn,secnumroman,tightenlines,amssymb,aps,nobibnotes,
  superscriptaddress,showpacs,balancelastpage,floatfix,nofootinbib]{revtex4-1}
\usepackage{graphicx,colordvi}
\usepackage{multirow}      
\usepackage{color}	
\usepackage[utf8]{inputenc}
\usepackage{amsmath}	
\expandafter\ifx\csname package@font\endcsname\relax\else
\expandafter\expandafter
\expandafter\usepackage
\expandafter{\csname package@font\endcsname}
\fi
\DeclareRobustCommand\substyle{\name@idx{document substyle}}
\DeclareRobustCommand\classoption{\name@idx{document class option}}
\DeclareRobustCommand\classname{\name@idx{document class}}
\def\name@idx#1#2{{\ttfamily#2}
\index{#2\space#1=\string\ttt{#2}\space#1}\index{#1>#2=\string\ttt{#2}}}

\def\d{\hbox{d}}
\def\be{\begin{equation}}
\def\ee{\end{equation}}
\def\bea{\begin{eqnarray}}
\def\eea{\end{eqnarray}}
\def\l{\label}

\def\exp{\hbox{exp}}

\def\siml{\hbox{\kern.1em \lower.6ex \hbox{$\sim$} \kern-1.12em
          \raise.6ex \hbox{$<$} \kern.1em }}
\def\simg{\hbox{\kern.1em \lower.6ex \hbox{$\sim$} \kern-1.12em
          \raise.6ex \hbox{$>$} \kern.1em }}

\begin{document}

\title{
      Paring correlations within the micro-macroscopic approach for the level
  density},

\author{A.G. Magner}
\email{Email: magner@kinr.kiev.ua}
\affiliation{\it  Institute for Nuclear Research, 03028 Kyiv, Ukraine}
\affiliation{\it Cyclotron Institute, Texas A\&M University,
  College Station, Texas 77843, USA}
\author{A.I. Sanzhur}
\affiliation{\it  Institute for Nuclear Research, 03028 Kyiv, Ukraine} 
\author{S.N. Fedotkin}
\affiliation{\it  Institute for Nuclear Research, 03028 Kyiv, Ukraine}
\author{A.I. Levon}
\affiliation{\it  Institute for Nuclear Research, 03028 Kyiv, Ukraine}
\author{U.V.~Grygoriev }
\affiliation{\it  Institute for Nuclear Research, 03028 Kyiv, Ukraine}
 \affiliation{University of Groningen,
9712 TS Groningen, Netherlands}
 
 \author{S. Shlomo}
\affiliation{\it Cyclotron Institute, Texas A\&M University,
  College Station, Texas 77843, USA}

\begin{abstract}
  Level density $\rho(E,N,Z)$ is calculated for the two-component
  close- and open-shell 
  nuclei
  with
  a given energy $E$, and neutron $N$ and proton $Z$ numbers,
  taking into account pairing effects
  within the microscopic-macroscopic approach (MMA).
       These analytical calculations
     have been carried out by using the semiclassical statistical
       mean-field approximations 
  beyond the saddle-point method of the Fermi gas model in a
   low excitation-energies range.  
     The level density $\rho$, obtained
       as function of the system entropy
  $S$, 
  depends essentially on the condensation energy $E_{\rm cond}$
  through the excitation energy
  $U$ in super-fluid nuclei. 
  The simplest super-fluid approach, based on the BCS theory,
  accounts for a smooth temperature dependence of the pairing gap $\Delta$
  due to particle number fluctuations.
  Taking into account the pairing  effects in
   magic or semi-magic nuclei, 
   excited below neutron
   resonances,
   one finds a notable pairing phase transition.
   Pairing correlations sometimes  improve 
      significantly
    the comparison with experimental data.
   
\date{August, 4nd, 2023}

\end{abstract}

\maketitle

\section{Introduction}

Many properties of heavy nuclei can be 
described
in terms of the level density
\cite{Er60,BM67,LLv5,Ig83,GF85,So90,Sh92,Ig98,AB03,EB09,AB16,KS18,ZV18,Ze19,KS20}.
   In close relation to
  the statistical level density \cite{Ig83,So90},  this article is devoted
  to the
    memory of Professor Peter Schuck, in particular, to his fruitful
    ideas of the
    semiclassical
pairing treatment in nuclear physics and associated areas
\cite{BS80,RS80,ER85,HR89plb,KR89zpa,FS00,VS03,BM04,BB05,VS10,VS11,PM13,SV13book,PS14,PS20,SU23}.
The level density 
$\rho(E,N,Z)$, where $E$,  $N$ and $Z$  
are the energy, neutron  and
  proton numbers,
respectively,
is given by
the inverse Laplace transformation of the partition function
$\mathcal{Z}(\beta,\boldsymbol\alpha)$ \cite{Er60}.
Within the grand
canonical ensemble 
the standard saddle-point method (SPM) is used
for integration
over all Lagrangian multipliers, $\beta$ for the energy $E$
and $\boldsymbol\alpha=\{\alpha_n,\alpha_p\}$ for neutron $N$ and proton $Z$
numbers \cite{Er60,BM67}.
This method assumes large
excitation energies $U$, so  that the temperature $T$ is
related to a  well-determined
saddle point $\beta^\ast$
in the integration variable
$\beta$ for a finite Fermi 
system of large neutron and proton numbers in a nucleus,
$T=1/\beta^\ast$. 
However,  data from
many experiments
 for energy levels and spins also exist for 
low excitation energy $U$,
where  such a saddle point does not exist. Moreover, there is a pairing
effect which can be extremely
important
at these $U$.
For presentation of experimental data on low-energy nuclear spectra,
the cumulative
level-density distribution 
-- cumulative number of quantum levels
below the excitation energy $U$ 
--
is conveniently often calculated 
for statistical analysis 
of
the experimental data on collective excitations \cite{Le20}. 
   Therefore, 
to simplify the calculations of the level density, $\rho(E,N,Z)$, 
 we 
 carry out
 \cite{KM79,MS21prc,MS21ijmpe,MS22npa,MS22ltp,MS23npae} the integration
 over the Lagrange multiplier
$\beta$ in the inverse Laplace
  transformation of the partition function
$\mathcal{Z}(\beta,\boldsymbol\alpha)$  
 analytically but more 
 accurately  beyond the SPM 
   for small and large shell-structure contributions \cite{BD72}.
  Thus, for 
the integration over $\beta$ we will use approximately the
micro-canonical ensemble 
which does not assume
a temperature and
an existence of thermodynamic equilibrium.

For formulation of the 
unified microscopic 
canonical 
and macroscopic grand-canonical 
approximation
 (MMA)   
to the level density \cite{KM79,MS21prc,MS21ijmpe,MS22npa,MS22ltp,MS23npae},
we found
a simple analytical
approximation for the
level density $\rho$ 
which satisfies
the two well-known limits.
One of them is the 
Fermi gas asymptote, 
$\rho \propto \exp(S)$ 
 for 
large entropy $S$ \cite{Er60}.  Another limit is 
the combinatoric expansion in powers of $S$ for a small entropy $S$
or excitation energy $U$;
see Refs.\ \cite{St58,Er60}.

    In the calculation of level density at low excitation energies in nuclei,
 we will consider  the system of interacting Fermi particles with a macroscopic
 number
 $A=N+Z$, described by the
 Hamiltonian of the well-known nuclear super-fluid model
 \cite{BMP58,Bel59}, 
 taking mainly the
 simplest Bardeen–Cooper–Schrieffer (BCS) \cite{BCS57} theory of
   superconductivity
 \cite{RS80,LLv9,ZV18,GF85}.
  This 
 method was used also in the
   level density calculations \cite{Ig83,So90} 
     for the description of the super-fluidity properties of nuclei, and
     for other 
     problems such as in nuclear astrophysics;
     see, e.g., Ref.~\cite{SC19}.
     We should emphasize the well known self-consistent method
      of
     super-fluidity
calculations,
the Hartree-Fock-Bogoliubov (HFB) theory \cite{RS80}.
Within a mean field approximation, as BCS or HFB
approaches,
 the pairing gap $\Delta$
depends sharply
on the
excitation energy $U$ in the phase transition from a super-fluid
to a normal
nuclear state. However,  as shown in Ref.~\cite{ER85}, if we take
into account also the particle number fluctuations by using the projected
HFB approach, the 
gap
$\Delta(T)$
becomes a smooth function of temperature $T$, 
  in contrast to any mean field 
  models.
  Smooth behavior of the pairing gap has been found also and
  carefully studied
  by analytical methods associated with the extended Thomas-Fermi
  (ETF) approach, see Refs.~\cite{BS80,HR89plb,KR89zpa,FS00,VS03,BM04,VS10}.
       Therefore, we cannot  use
the normal-state properties of 
the shell closure, in particular, for
magic nuclei like
$^{40}$Ca, $^{48}$Ca, $^{56}$Ni, and $^{208}$Pb, to
deduce that the pairing transition from a super-fluidity to the normal
state does not exist there. Our present work was
partially initiated also by the recent experimental studies 
in
Ref.~\cite{BC22} in order to clarify their relation to the super-fluidity in
magic nuclei.
Pairing correlation can in fact
influences the level density in magic nuclei, and the question is what are
other reasons for difficulty in its experimental observation
 in magic nuclei, see Refs.\ \cite{RS80,MS23npae}. This question
cannot be
separated from the study of the shell structure effects in magic nuclei, and
it is the main question
under consideration in this work.

For a deeper understanding of the 
correspondence between the classical and the quantum approach and 
  simplifying
the
problem 
for analytical derivations,
it is worthwhile to analyze the shell and pairing
effects in the statistically averaged level density $\rho$,
see Refs.~\cite{KM79,Ig83,So90}, 
by using the semiclassical
periodic-orbit (PO) theory (POT) 
\cite{SM76,BB03,MY11}.
We 
extended the MMA
approach \cite{KM79} in
Refs.~\cite{MS21prc,MS21ijmpe,MS22npa,MS22ltp,MS23npae}, 
for
semiclassical description of the shell and isotopically
asymmetry 
effects in the level density of complex nuclei.
Smooth properties of the level density  
as function of the nucleon number $A$
have been studied
within the framework of 
self-consistent
ETF approach \cite{BG85,Sh92,KS18}.
However, for instance, the 
pairing effects in the statistical
MMA level density $\rho$
are still 
attractive subjects
\cite{Ig83,So90,BD72,ZV18,Ze19}. 
    See also common ideas and a very intensive recent analytical
study of pairing effects within the  semiclassical Wigner-Kirkwood  $\hbar$
expansion 
including nuclear surface $\hbar^2$ corrections; see
Refs. \cite{SV13book,SU23}, and
references therein.

In the present paper,  we  shall first present 
the MMA results for the shell-dependent level
density $\rho(E,N,Z)$ taking into account pairing correlations through the
condensation energy shifts for
magic (close-shells) $^{40}$Ca, $^{48}$Ca, 
$^{56}$Ni and $^{208}$Pb, and non-magic (open-shells)
$^{52}$Fe nuclei, and several semi-magic nuclei.
Some of these symmetric nuclei, $^{56}$Ni and $^{52}$Fe, were
studied recently
experimentally in Ref.~\cite{BC22}. 
    Then, several other situations
for isotopically asymmetric nuclei were
presented to show pairing effects. 
Here we concentrate on
low energy states (LES) of
nuclear excitation-energy spectra below
neutron resonances in  magic isotopically symmetrical
    and asymmetric
nuclei, and
 non-magic one for comparison. 

\section{Microscopic-macroscopic approach}

For 
the statistical
description of level density $\rho$ of a nucleus in
  terms of 
  the total energy, $E$, and the neutron, $N$, and proton, $Z$, numbers,
 one
can begin with
the micro-canonical expression \cite{Er60,BM67}:
\be
\rho(E,N,Z)=
\int \frac{\d \beta \d \boldsymbol\alpha}{(2\pi i)^3}~
\exp\left[S\left(\beta,\boldsymbol\alpha\right)\right], \l{dendef}
\ee
where 
$S=\ln \mathcal{Z}(\beta,\boldsymbol\alpha)
+\beta E 
-\boldsymbol\alpha {\bf N}$, and
${\bf N}=\{N,Z\}$
with the total particle number $A=N+Z$.
The partition function $\mathcal{Z}$ depends on the Lagrange
multipliers, $\beta$ and
$\boldsymbol\alpha=\{\alpha_n,\alpha_p \}=\beta \boldsymbol\lambda$. 
The neutron and proton
chemical potential components are given by
$\boldsymbol\lambda=\{\lambda_n,\lambda_p\}$,
where $\lambda_n=\alpha_n/\beta $ and $\lambda_p=\alpha_p/\beta $.
The entropy $S(\beta,\boldsymbol\alpha)$ can be expanded in power series
over $\boldsymbol\alpha$ for a given $\beta$ near the saddle point
$\boldsymbol\alpha^\ast$,
\bea\l{Sexp}
&S(\beta,\boldsymbol\alpha)=S(\beta,\boldsymbol\alpha^\ast)\nonumber\\
&+(1/2) \left(\partial^2 S/\partial \boldsymbol\alpha^2\right)^\ast
\left(\boldsymbol\alpha-\boldsymbol\alpha^\ast\right)^2+\ldots~.
\eea
The Lagrange multiplier, $\boldsymbol\alpha^\ast$, and the chemical potential
$\boldsymbol\lambda$,
are defined in terms of the neutron and proton relatively
    large
particle numbers
${\bf N}$ by a 
saddle-point condition,
\begin{equation}\label{Seqsd}
  \left(\frac{\partial S}{\partial \boldsymbol\alpha}\right)_{\alpha=\alpha^\ast}
  \equiv
  \left(\frac{\partial \ln \mathcal{Z}}{\partial
    \boldsymbol\alpha}\right)_{\alpha=\alpha^\ast}-{\bf N}
=0~.
\end{equation}
In order to avoid the SPM divergences for the zero
excitation energy limit,
  we will integrate over $\beta$ in Eq.~(\ref{dendef}) more accurately using
  the statistically averaged excitation energy  $U$ as a 
  measure of the mean nuclear excitations, instead of the
  temperature $T$. Then, $\beta $ will be the integration variable which
  practically coincides with the saddle point, $\beta^\ast=1/T$,
      for
  sufficiently large
  excitation energies. In particular, one can 
      find \cite{LLv9} the critical value
  of the temperature
  $T_{\rm c}$ for disappearance of pairing correlations.
Introducing, for convenience, the potential
$\Omega=-\mbox{ln}\mathcal{Z}/\beta$ in the mean field approach, one has to
specify the system through the Hamiltonian 
taking into account the pairing correlations within the
    simplest approach
based on the BCS
 approach \cite{BCS57,RS80,LLv9}.

 The statistically averaged condensation
 energy $E_{\rm cond}$ can be derived 
  in simple form in terms of
 the constant
pairing gap $\Delta$, independent of the quasi-particle spectrum
\cite{BM67,RS80,SC19}.
For constant $\Delta \approx \Delta_0$, one can use its
averaged empiric dependence on the particle number $A$,
$\Delta_0\approx 12 A^{-1/2}$  MeV
 \cite{BM67,RS80,Ig83,So90}.
 The results for the level density are weakly 
dependent on the variation (within 15\%) of the 
number in front of $A^{-1/2}$, also when we replace this power dependence by
that suggested in   
Ref.~\cite{Vo84}.
This phenomenological behavior $\Delta(A)$ 
    is 
good for sufficiently heavy
nuclei, in particular, for $A \simg 40$.

Following Refs.~\cite{Ig83,So90}, one can reduce
the level density calculation for the system
of interacting Fermi-particles described by the two-body Hamiltonian to
that of the system of a mean field with the
quasi-particles Hamiltonian \cite{RS80,GF85}.
For $N$ neutrons ($Z$ protons)
one writes
\bea\l{H}
&\hat{H} - \lambda \hat{N}=
\sum_{js}\left(\varepsilon_{j}-\lambda\right)\hat{a}_{js}^+\hat{a}_{js}
  \nonumber\\
&- G 
\sum_{jj^{\prime}}
\hat{a}^+_{j+} \hat{a}^{+}_{j-}
\hat{a}_{j^{\prime}-}\hat{a}_{j^\prime+}~.
\eea
Here, $\varepsilon_j$ is the single-particle energy of states $j$,
which are doubly 
degenerated over the spin projection sign, $s=\pm$, and
$\lambda \approx \lambda_n\approx \lambda_p$
is the chemical potential.
 In Eq.~(\ref{H}), $a^+_{js}$ and $a_{js}$ are the operators of
the creation and
annihilation  of particles, 
respectively; see
details in Ref.~\cite{So90}.
The second term in
Eq.~(\ref{H}) is the pairing interaction with the constant $G$
($G\approx G_n\approx G_p$),
  which is the averaged matrix element of the residue interaction.
  In this case
  one has a very simple and powerful model for the description of the
  pairing properties of nuclei.  For the
  operator of the total particle number, $\hat{N}$, one has
$\hat{N}=\sum_{js} \hat{a}^+_{js}\hat{a}_{js}$~.
The interaction constant
$G$ (neutron $G_n$ and proton  $G_p$) can be
determined using 
experimental data
\cite{BM67,So90}. 
The corresponding thermodynamic averages of any operator $\hat{Q}$
are determined
by
$\langle \hat{Q} \rangle =
\mbox{Tr} \left[\hat{Q}\mbox{exp}\left(-\beta H\right)\right]/
\mbox{Tr} \left[\mbox{exp}\left(-\beta H\right)\right]$.
Up to a constant, the Hamiltonian $H$, Eq.~(\ref{H}), coincides
with that of
the 
Fermi-quasi-particles in a mean field. Therefore,
for the entropy $S$, one can use
the similar expression:
$S=2\sum_j\left[\beta \epsilon_j \overline{n}_j -
  \ln\left(1-\overline{n}_j\right)\right]$,
where
$\epsilon_j=[(\varepsilon_j-\lambda)^2+\Delta^2]^{1/2}$,
and $\overline{n}_j=\left[1+\exp\left(\beta \epsilon_j\right)\right]^{-1}$
are the quasi-particle energies, and
occupation numbers averages, respectively \cite{Ig83}.
Straightforward 
    analytical derivations \cite{LLv9,Ig83,GF85,So90} valid near
the critical point of the superfluid-normal phase transition
lead to the
critical temperature $T_{\rm c}$,
\be\l{Tc}
T_{\rm c}=e^C \Delta_0/\pi~,
\ee
where $C\approx 0.577$ is the Euler constant.

For a given temperature $T$, 
when exists, by minimization of the expectation
value of the grand-canonical potential $\Omega$, one has
(see Refs.~\cite{Ig83,So90}),
$\Omega\equiv E-\boldsymbol\lambda 
-{\bf N}-S/\beta
= \langle \hat{H} - \boldsymbol\lambda 
\hat{{\bf N}}
- \hat{S}/\beta\rangle~$,
where  $\langle ... \rangle$ denotes a statistical average 
over the operator
enclosed in angle brackets. 
Here,
$\hat{{\bf N}}$
is the
particle (neutron and proton) number and $\hat{S}$ is
the entropy operators.
For the pairing ground-state energy
$\langle H_0 \rangle$, which equals
$\langle H\rangle$ at zero excitation energy,
$U=0$, one finds
$\langle H_{0} \rangle \approx \Delta_0^2/4G$. 
With the heat part, $U_c=a T_c^2$ [Eq.~(\ref{Tc}) for $T_c$], where $a$ is
the level density parameter,  one obtains
for the total excitation energy
$U^{\rm tot}_c$ of the mean superfluid-collapse transition,
\be\l{Utotc}
U^{\rm tot}_{\rm c}=a T_c^2 + \Delta_0^2/4G~.
\ee

Following Refs.~\cite{MS21prc,MS21ijmpe},
one can obtain analytical expressions for the level density
$\rho(E,N,Z)$, beyond the standard SPM, by
calculating more accurately the
integral over $\beta$ of the inverse Laplace presentation (\ref{dendef}),
    \begin{equation}\label{denbesnp}
    \rho \approx \rho^{}_{\tt{MMA}}(S)
    =\overline{\rho}_\nu S^{-\nu}I_{\nu}(S)~,
    \end{equation}
    where $I_{\nu}(S)$ is
    the modified Bessel function of the index $\nu$. 
        In this function,
    one has $\nu=3$ for the shell
   structure dominance and $\nu=2$ for small shell contributions.
   For the entropy $S$, one finds
   \be\l{Seff}
S\approx S_{\rm eff}=2 \sqrt{a U_{\rm eff}}~,
\ee
where $a$ is the level density parameter, and
$U_{\rm eff}$ is the excitation energy, 
    shifted due to the pairing correlations,
\be\l{Ueff}
U_{\rm eff}=U-E_{\rm cond}~\ge 0.
\ee
   For the condensation energy $E_{\rm cond}$,
      one finally has
      \cite{Ig83,RS80,So90}
\be\l{Ec}
E_{\rm cond}=\frac{3 a \Delta_0^2}{2\pi^2}
\approx \frac{216}{\pi^2 K} ~,
\ee
where 
$K=A/a$ is 
the inverse level density parameter. For $K\sim 10-30$ MeV
\cite{KS18,MS21prc,MS21ijmpe}, one obtains $E_{\rm cond}\approx 1-2$ MeV.
Notice also that, the condensation energy
$E_{\rm cond}$, Eq.~(\ref{Ec}),  depends
on the particle number $A$ mainly
through the inverse level density
parameter K for $\Delta_0=12/A^{1/2}$ MeV. This parameter 
  depends on $A$ \cite{MS21prc,MS21ijmpe},
basically through
shell effects.
For large and small $S$, one obtains from the general
 equation (\ref{denbesnp}) 
the well known Fermi gas \cite{Er60} and the
combinatoric \cite{St58}
asymptotes using 
the important $1/S$ and $S^2$ corrections, respectively.
     
    In Eq.~(\ref{denbesnp}), the value of $\nu$ in Eq.~(\ref{denbesnp}) depends
     on the number of the integrals of motion and of the shell structure
     contribution, which is 
     determined by
\be\l{pars}
\xi
\approx\frac{8\pi^6 U_{\rm eff} A^{1/3}}{3a\lambda^2}\mathcal{E}_{\rm sh}~,
\quad \mathcal{E}_{\rm sh}=-\frac{\delta E}{E_{\rm ETF}}~A~,
\ee
where 
$\mathcal{E}_{\rm sh}$ is the relative energy shell correction, and
$\delta E$ is in units of the smooth 
    ETF energy, $E_{\rm ETF}$, per particle \cite{BG85,KS20,MS21prc}.
   The energy shell correction, $\delta E$,
  can be approximated,
  for a major shell structure,  within
  the semiclassical POT accuracy
  (see Refs.~\cite{SM76,BB03,MY11,MS21prc,MS21ijmpe,MS23npae}). 
  The characteristic parameter $\xi$, Eq.~(\ref{pars}), proportional to
   $\mathcal{E}_{\rm sh}$, specifies
  the two different approximations,  $\xi \ll 1$ and  $\xi \gg 1$,
  for small  and large shell correction
  contribution $\mathcal{E}_{\rm sh}$, respectively.
  We will consider below mainly nuclei,
  for which one has the case of relatively large $\mathcal{E}_{\rm sh}$ and,
  therefore, large
  $\xi$. 
   As shown in Ref.~\cite{MS21ijmpe}, one thus finds 
  that $\nu=3$ and $\overline{\rho}^{}_3
   \propto a^2/\sqrt{\overline{\xi}}$, where
   $\overline{\xi}=A^{1/3}\mathcal{E}_{\rm sh}/\lambda^2$
   in Eq.~(\ref{denbesnp}) for the MMA2 approach while $\nu=2$ and
   $\overline{\rho}^{}_2\propto a$  for small $\xi$ in the MMA1 approach.
   Within the  MMA2 approach, one 
   may also use an
   analytical
   approach for small values of shell corrections
   but with large shell contributions
       due to their significant derivatives.
     The reason is
       large derivatives of the grand-canonical potential
       $\Omega$  over the chemical potential $\lambda$, MMA2b 
           approach.
     This is in contrast to the
     MMA2a approach with evaluations of the  $\mathcal{E}_{\rm sh}$ by using the
     numerical results of Ref.~\cite{MSIS12}.
   The SPM asymptotes FG at $\xi \rightarrow 0$,
  Refs.~\cite{Er60,MS21ijmpe}, 
are divergent
at $U_{\rm eff}\rightarrow 0$.

\section{Discussion}

\begin{figure*}
  \vskip1mm
    \includegraphics[width=14.0cm]{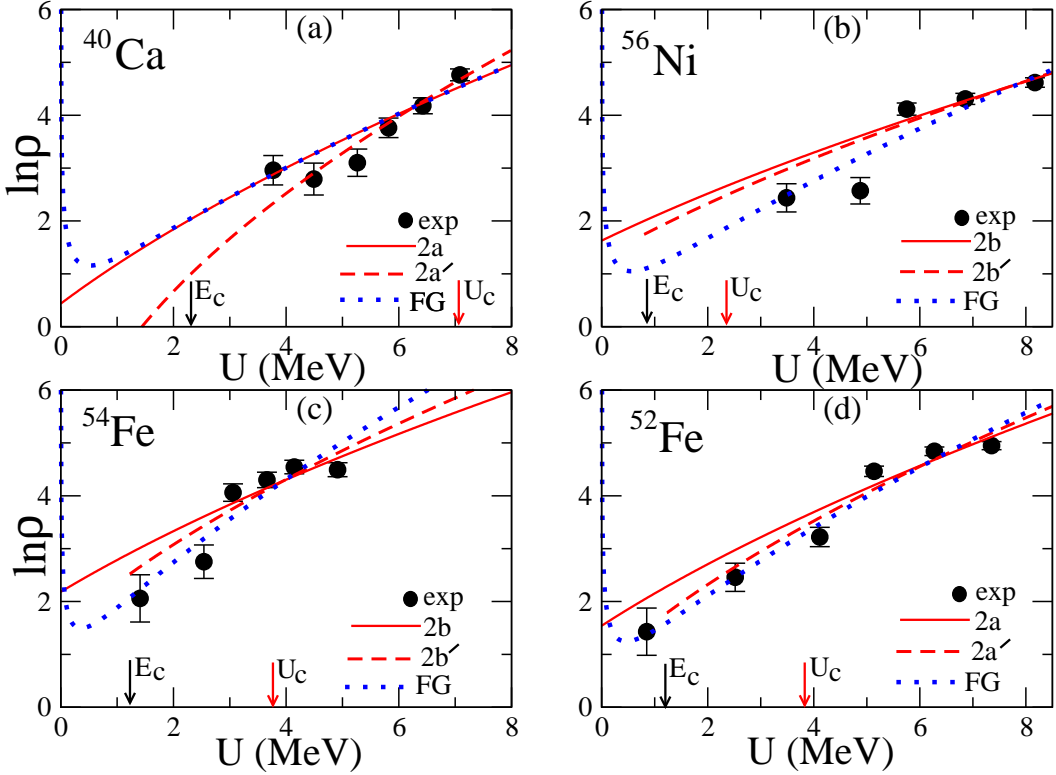}

  \vskip12mm\caption{{\small 
       Level density, $\mbox{ln}\rho(E,N,Z)$,
     for
     low energy states in the 
     magic (close-shell) $^{40}$Ca (a) and $^{56}$Ni (b),
     semi-magic $^{54}$Fe (c)
     and non-magic (open-shell) $^{52}$Fe (d) nuclei.
       Solid lines ``1'' show the MMA approach for the minimal value of
    the LMS error $\sigma$,
     Eq.~(\ref{denbesnp}), Table~\ref{table-1}, by
     neglecting the pairing condensation. Dashed lines ``2'' are the
    same but taking
    into account
    the pairing effect
    through the condensation energy
    $E_{\rm cond}$ [Table \ref{table-1}, Eq.~(\ref{Ec})].
       Blue dotted lines ``3'' present
       the Fermi gas (FG) approach \cite{Er60}. 
           The chemical potential is
      $\lambda=40$ MeV.
       Experimental close circles     
     are obtained directly from the 
     excitation energy data \cite{ENSDFdatabase}
    by using the sample method \cite{MS21prc,MS21ijmpe}.
    The plateau condition was checked
    within the number of dots 
    $\aleph=5-8$
    over the inverse
    level density parameter
    $K$.  Black ($E_{c}=E_{\rm cond}$), and
    red ($U_c=U^{\rm tot}_{c}$) arrows
    present
    the evaluations of the condensation energy (\ref{Ec}), and the
    excitation energy (\ref{Utotc}) for the phase superfluid-normal transition,
     respectively.
}}
\label{fig1}
\end{figure*}

\begin{figure*}
  \vskip8mm
    \includegraphics[width=14.0cm]{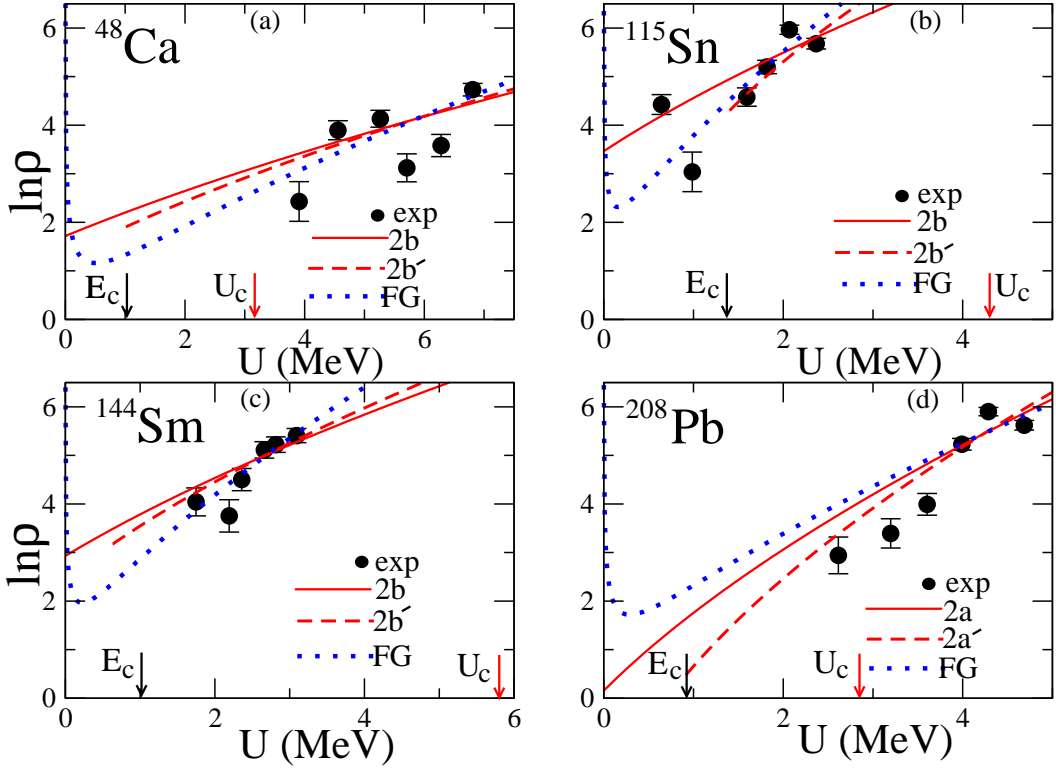}

  \vskip12mm\caption{{\small 
      The same as in Fig.~\ref{fig1} but for other nuclei $^{48}$Ca,
      $^{115}$Sn, $^{144}$Sm, and $^{208}$Pb.
}}
\label{fig2}
\end{figure*}

In Figs.\ \ref{fig1} (Table~\ref{table-1}) and \ref{fig2}
 (Table~\ref{table-2}) 
 we present results of
 theoretical calculations of   
 the statistical level density $\rho(E,N,Z)$ (in logarithms) within the MMA
 approach, 
Eq.~(\ref{denbesnp}), and its FG limit \cite{Er60},
as functions of the
excitation energy $U$ for different nuclei.
    In Fig.1, these results are compared 
with the experimental
data for the close-shell magic
    nuclei $^{40}$Ca (a) and  $^{56}$Ni (b)
  semi-magic nucleus $^{54}$Fe (c)
and open-shell non-magic 
   nucleus
  $^{52}$Fe (d),
  accounting for pairing 
    contributions.
  Pairing effects in $^{56}$Ni and $^{52}$Fe nuclei
  are studied experimentally in details in Ref.~\cite{BC22}. 
      In Fig.~\ref{fig2}
  we display
  other results for $^{48}$Ca, $^{115}$Sn,
   $^{144}$Sm, and $^{208}$Pb nuclei.
Our results 
are obtained  by using 
the values of the inverse level density parameter $K$,  found
from their one-parametric least mean-square 
fits  (LMSF) of the MMA results
to the experimental  data \cite{ENSDFdatabase}.
 The 
 data shown by dots with error bars in 
     Figs.~\ref{fig1} and \ref{fig2}
are obtained for the statistical level density $\rho(E,N,Z)$
from
the experimental ENSDF \cite{ENSDFdatabase} data
on the excitation energies $U$ and spins $I$ of the
quantum states spectra
by using the sample method,  
    $\rho_i=L_i/U_s$, where $L_i$
is the number of states in the $i$-th sample, 
$i=1,2,...,\aleph$
\cite{So90,LLv5,MS21prc}. 
    The dots are plotted at  mean-weighted positions
$U_i$ of the excitation energies for each $i$th sample.
Convergence of the sample method over the equivalent sample-length
parameter $U_{s}$ of the statistical averaging 
was studied 
    under the 
    statistical plateau condition over the $K$ values. 
    The lengths $U_s$ (or the equivalent number of samples 
    $\aleph$) play  
a role which is similar to that of 
averaging parameters in the 
Strutinsky smoothing procedure 
for the shell correction method (SCM)
calculations of the averaged single-particle
level density
\cite{BD72}.
The plateau  condition in our calculations means almost constant 
value of the 
physical parameter $K$ (with better than 20\% 
accuracy)
within a relatively
 large change of
 sample numbers 
 $\aleph$.
Therefore, the results
of Tables
\ref{table-1} and \ref{table-2},
calculated at the same  values of the found plateau, 
 do not depend, 
with the statistical accuracy, on the averaging
parameter 
$\aleph$
within the plateau.
 The {\it statistical condition,} $L_i\gg 1$, at different 
 plateau values 
 $\aleph$
 determines the accuracy
    of our statistical calculations. 
    As in the SCM, in our calculations
    within the sample method  
    with  good plateau values for the sample numbers, 
$\aleph$,
    shown 
       in the caption of 
        Fig.~\ref{fig1},  
            one obtains a
            relatively smooth 
            statistical level density as function of the excitation energy $U$.
               We 
 require such a smooth behavior  because 
 the statistical fluctuations 
 of particle numbers are neglected 
in our theoretical derivations of the level density.
\begin{table}[pt]
\begin{tabular}{|c|c|c|c|c|c|c|c|c|c|c|c|c|c|c|c|}
\hline
Nuclei & MMA &$\mathcal{E}_{\rm sh}$
& $E_{\rm cond}$ 
&$U^{\rm tot}_{\rm c}$& $K$& $\Delta K$ &
$\sigma$  \\
 &  & & (MeV)& (MeV) &(MeV) & (MeV)&  
\\
\hline
$^{40}$Ca & 2a$^\prime$ &0.061 & 2.3 & 7.1  & 9.6 & 0.3  & 1.3 \\
\hline
& 2a & &  &   & 12.5 & 0.3  & 1.6 \\
\hline
 & FG & &  &   & 13.4 & 0.4  & 1.5 \\
\hline
$^{56}$Ni & 2b$^\prime$ &0.008 & 0.80 &2.50  & 27.3 & 0.7  & 2.2 \\
\hline
& 2b & &  &   & 29.0 & 0.7  & 2.4 \\
\hline
& FG & &  &   & 17.8 & 0.7  & 3.2 \\
\hline
$^{54}$Fe & 2b$^\prime$ &0.264 & 1.23 & 3.82  &  17.9 & 0.8  & 1.9 \\
\hline
& 2b & &  &   &  21.1 & 0.7  & 1.9 \\
\hline
 & FG & &  &   &  12.5 & 0.5  & 2.3 \\
\hline
$^{52}$Fe & 2a$^\prime$ &0.003 & 1.23 & 3.84  &  17.7 & 0.5  & 2.5 \\
\hline
& 2a & &  &   &  20.1 & 0.5  & 2.4 \\
\hline
 & FG & &  &   &  15.4 & 0.5  & 2.7 \\
\hline
\end{tabular}

\vspace{-0.2cm}
\caption{{\small
    The inverse level density parameter $K$ (sixth column) and its error
    (seventh column),
  found by the LMSF with 
    relative accuracy $\sigma$ (eighth column),
    for the nuclei shown in the first column (see Fig.~\ref{fig1}).
    The second column displays the MMA method:
    MMA2a or MMA2b with no pairing, $E_{\rm cond}=0$, and MMA2a$^\prime$ or
      MMA2b$^\prime$ with the finite condensation energy $E_{\rm cond}$, all
      versus FG model. The
    relative shell (with pairing)
    corrections $\mathcal{E}_{\rm sh}$,
    Eq.~(\ref{pars}) (from Ref.~\cite{MSIS12}), are shown
      in the third column.
    The fourth and fifth columns
    show the found
    condensation energies $E_{\rm cond}$, Eq.~(\ref{Ec}), and
    total excitation energies $U^{\rm tot}_{\rm c}$ for the phase transition
    from superconductive to normal nuclear states,
    Eq.~(\ref{Utotc}), respectively. The experimental excitation spectra
    were taken from
    the ENDSF base
    \cite{ENSDFdatabase} as in Fig.~\ref{fig1};  see the text.
}}
\label{table-1}
\end{table}
\begin{table}[pt]
\begin{tabular}{|c|c|c|c|c|c|c|c|c|c|c|c|c|c|c|c|}
\hline
Nuclei & MMA &$\mathcal{E}_{\rm sh}$
& $E_{\rm cond}$ 
&$U^{\rm tot}_{\rm c}$& $K$& $\Delta K$ &
$\sigma$  \\
 &  & & (MeV)& (MeV) &(MeV) & (MeV)&  
\\
\hline
$^{48}$Ca & 2b$^\prime$ &0.224 & 1.0 & 3.2  & 21.6 & 1.3  & 2.6 \\
\hline
& 2b & &  &   & 23.7 & 1.2  & 2.6 \\
\hline
 & FG & &  &   & 15.5 & 0.8  & 2.6 \\
\hline
$^{115}$Sn & 2b$^\prime$ & 0.061& 1.4 &4.4  & 15.6& 1.7 & 3.0 \\
\hline
& 2b & &  &   & 23.7 & 1.1  & 3.1 \\
\hline
 & FG & &  &   & 11.7 & 0.6  & 3.5\\
\hline
$^{144}$Sm & 2b$^\prime$ & 0.368& 0.6 &2.0  &  34.2 &1.2  & 1.4 \\
\hline
& 2b & &  &   &  38.8 & 1.3  & 1.6 \\
\hline
 & FG & &  &   &  20.8 & 0.4  & 1.1 \\
\hline
$^{208}$Pb & 2a$^\prime$ &1.77 & 0.9 & 2.9   & 23.9 & 0.9  & 3.2 \\
\hline
& 2a & &  &    & 28.1 & 0.9  & 3.4 \\
\hline
 & FG & &  &    & 38.3 & 1.6  & 3.6 \\
\hline
\end{tabular}

\vspace{-0.2cm}
\caption{{\small
    The same as in Table~\ref{table-1} but for nuclei $^{48}$Ca, $^{115}$Sn,
   $^{144}$Sm, and $^{208}$Pb, presented in Fig.~\ref{fig2}. 
}}
\label{table-2}
\end{table}
The relative error $\sigma$ of the standard LMSF
(see 
Tables \ref{table-1} and  \ref{table-2}), 
for the description of the spectra
data, in terms of the
statistically averaged
level density
$\rho_i$, 
is given by the standard formula through the $\chi^2$
of the  LMSF,
$\sigma^2=\chi^2/(\aleph-1)$ is the error dispersion.
The error $\sigma$
determines the applicability 
of
the theoretical 
approximations, $\rho(U_i)$.
These experimental results are practically independent of model because
of the using of the
plateau condition over the number of points 
$\aleph$
(Ref.~\cite{BD72}).
We do not use 
empiric (nonphysical) free fitting parameters.
 We discuss the level density $\rho(E,N,Z)$ integrated over spins 
   accounting for
 degeneracy over their projections.
In particular, this $\rho(E,N,Z)$  is independent of assumptions for using
the approximation of small spins,
        and there is no explicit dependence of
  $\rho$ on the
nuclear moment of inertia.

Fig.~\ref{fig1} 
shows the results for
 four different situations concerning pairing
contributions. In the upper plots of Fig.~\ref{fig1}
we present two magic nuclei
with the red pairing-collapse arrow far away on right of the 1st
dot for $^{40}$Ca (a), and slightly on its left
 for $^{56}$Ni (b), respectively.
In both panels,
one has a significant excitation-energy gap below the first
point in close-shell magic nuclei. In contrast to
them, in the lower plots Fig.~\ref{fig1} (c) and (d) we show a
semi-magic and non-magic  
  examples for the nuclei $^{54}$Fe and $^{52}$Fe, respectively, where
the first point appears at relatively small excitation
energy $U$.  The
super-fluid collapse (red arrow) is placed in 
    Fig.~\ref{fig1} (c,d) on a
finite distance from the
first excited state, which is of the order of the condensation
energy $E_{\rm cond}$.

Thus, we do not expect significant pairing effects for the nucleus $^{56}$Ni
(Fig.~\ref{fig1} (b) and Table~\ref{table-1}) 
which agrees with experimental studies in Ref.~\cite{BC22}.
However, it is related to the superfluid-normal phase
transition itself,
rather than to a 
shell closure in nucleus
$^{56}$Ni 
in its normal state.
Best conditions for the experimental observation of a
super-fluidity transition 
in studied nuclei can be found, in our opinion, in $^{40}$Ca,
see Fig.~\ref{fig1} (a)
and Table \ref{table-1} because of sufficiently large
distance between the
first excited state and estimated
$U_c^{\rm tot}$, Eq.~(\ref{Utotc}), for
the pairing collapse. 
    It is seen also from a significant difference of the MMA
results ``2'' for the finite condensation energy $E_{\rm cond}$ and those ``1'' at
$E_{\rm cond}=0$.
An intermediate situation for pairing observation takes place in
 Fig.~\ref{fig1} (c,d)
for a semi-magic $^{54}$Fe and non-magic $^{52}$Fe,
see also
 Table \ref{table-1}.

In all plots of Fig.~\ref{fig1}  and Table~\ref{table-1}
we present the
dominating shell-structure
MMA approach
(\ref{denbesnp}) with large $\xi$, 
    especially large for  $^{54}$Fe
[Fig.~\ref{fig1} (c)].
  The MMA2a approach \cite{MS21ijmpe} with the relative
  shell-correction energy
  $\mathcal{E}_{\rm sh}$
  from Ref.~\cite{MSIS12}  is associated with a smallest LMSF
  error $\sigma$
  for calculations in
  Fig.~\ref{fig1}.

Thus,
the MMA2a results presented in Fig.~\ref{fig1} (a) for $^{40}$Ca
show a strong shell and
pairing effects with relatively large
$\mathcal{E}_{\rm sh}$ ($\xi \gg 1$); see
also Table \ref{table-1} and Ref.~\cite{MSIS12}.
The pairing energy-condensation contributions [cf. dashed ``2'' and solid
``1'' lines in
Fig.~\ref{fig1}] are significant,
especially remarkably seen in  Fig.~\ref{fig1} (a) for $^{40}$Ca.
They notably improve
  the comparison between the
experimental dots and theoretical MMA dashed results.
These should be 
compared with those of the pairing 
 discard version (``1'')
of the MMA
approach [MMA2a in (a,d) and MMA2b in (b,c)] shown by
solid lines.  The pairing effects 
are 
smaller 
in Fig.~\ref{fig1} (d) and, especially, in
 panels (b,c)  than in that of (a).

Fig.~2 (Table~\ref{table-2}) shows several other situations concerning
the pairing  contributions.
In Fig.~2 (a) we present the same level density as in Fig.~\ref{fig1} (a) but
for another double magic isotope $^{48}$Ca with a large isotopic asymmetry
parameter, $X\approx 0.2$. There is a super-fluidity but essentially below
all experimental points as shown by the red arrow.
Therefore, the pairing correlations are not easily
detected, in contrast to those for  $^{40}$Ca [Fig.~\ref{fig1} (a)]. The error
parameter $\sigma$ and the value of the inverse level parameter $K$ are
significantly high for $^{48}$Ca than in the case of $^{40}$Ca.
Relatively, the super-fluidity effect measured by the difference between
dashed and solid red curves
 decreases significantly for $^{48}$Ca.

The next plot [Fig.~\ref{fig2} (b)] for a semi-magic even-odd nucleus
$^{115}$Sn is devoted to another exotic case concerning super-fluidity.
Notice that in this case there are several excitation energies below
the condensation energy (two lowest points).
This is because there are several lowest nuclear excited states which are
significantly below the condensation energy value $E_{\rm cond}$.
 These excited states are excluded from our calculations
because they give relatively very small contribution to the error
dispersion $\sigma^2$
(about 4\%), and almost do not change the result 
(see Table~\ref{table-2})
for the dispersion.
Therefore, in this case, almost all experimental
points, giving  important contributions into the
dispersion parameter $\sigma^2$, are below the superfluidity-normal phase
transition evaluated by the red arrow. Thus, one can assume that the even-odd
nucleus $^{115}$Sn is in principle superfluity, as seen also from essential
difference of curves ``2'' and ``1''.
In contrast to this, within the standard Fermi-gas model, one has
negligibly small
pairing effects, due to almost zero corresponding change of the
background energy in this FG model \cite{BM67}.

In the next plot (c) of Fig.~\ref{fig2} for another semi-magic nucleus
$^{144}$Sm, all experimental points are lying inside of the interval between
both arrows $E_c$ and $U_c$, that perhaps leads to a good condition for
detection of the super-fluidity effects. However, the difference between
solid and dashed curves, which is a measure of the superfluity effect,
is small in this case, in contrast to the super-fluidity of magic nucleus
$^{40}$Ca [Fig.~\ref{fig1} (a)] 
    but similar to that for another semi-magic
nucleus $^{54}$Fe [Fig.~\ref{fig1} (c)].

Much larger difference occurs
in the case (d) of Fig.~\ref{fig2} for
double magic nucleus $^{208}$Pb with about the same asymmetry parameter
$X\approx 0.2$ as in the panel (a) of this figure
for another double magic nucleus $^{48}$Ca [Fig.~\ref{fig2} (a)]. But 
a similar problem for experimental observation of the superfluity effects
takes place as in 
 Fig.~\ref{fig2} (a)
for $^{48}$Ca, and even more similar to those for $^{56}$Ni in
Fig.~\ref{fig1} (b). The distance of the pairing collapse energy
$U_c$ from the first
excitation energy state point is too small
    in Fig.~\ref{fig2} (d), almost 300 keV but on its right.
Therefore, it is difficult
to detect the super-fluidity effects in $^{208}$Pb.
 In close similarity with 
    situation for magic nuclei $^{56}$Ni and  $^{48}$Ca:
    There is a superfluity transition to the
normal state but it is very difficult or even
impossible to observe because
of a small distance of the pairing collapse energy
$U^{\rm tot}_c$, Eq.~(\ref{Utotc}), to the first
excitation energy point.

As shown in Tables \ref{table-1} and  \ref{table-2},
the values of the inverse level density
parameter K can be significantly larger than those (about 10 MeV) for the
neutron resonances.
 In all plots of Figs.~\ref{fig1} and \ref{fig2}, 
one can clearly see also the divergence
of
the SPM FG lines ``3'' \cite{Er60}
near the zero excitation energy,
$U\rightarrow 0$. 
 The FG approach has larger (or sometimes, of the order)
 value of $\sigma$ for considered low-energy states in all plots of 
 Figs.~\ref{fig1} and \ref{fig2}, except for that of the semi-magic
 nucleus  $^{144}$Sm and magic nucleus $^{48}$Ca.

\section{Conclusions}

We have obtained agreement between the results of the
 theoretical approximation (MMA) and experimental data
  \cite{ENSDFdatabase} for  the
statistical level density
$\rho$  in the low excitation energy states (LES)
region for several nuclei with close and open shells and for
intermediate situations. Using
the mixed micro-  and grand-canonical ensembles beyond
the standard saddle-point method of the Fermi gas model we take into
account the pairing correlations.
 The derived level density expressions
 can be 
 approximated by those known as small (combinatoric) and relatively
 large (Fermi gas) limits with
 excitation energies
 $U_{\rm eff}$, shifted 
 due to pairing correlations.
The MMA approach
    clearly manifests an 
advantage
over the standard full SPM Fermi gas (FG) approaches
at low excitation energies, because 
 it does not diverge
in the limit of
small
excitation energies. 
The values of the inverse
level density
parameter $K$ 
obtained  within the MMA approach
for LES's below
neutron resonances in 
spectra of 
several super-fluid nuclei are found to be
    smaller than those obtained
at zero condensation energy $E_{\rm cond}$.
The pairing correlation contributions to the MMA
approach significantly improve agreement with experimental data for
magic nuclei as $^{40}$Ca.
An existing 
opinion 
on the absence
of pairing effects in
magic (close-shell) nuclei as $^{208}$Pb and $^{56}$Ni 
  can be explained
by a very
short spectrum length
within 
the super-fluid phase transition. Therefore,  
    it might be difficult 
    to detect such effects.
As shown for another magic nucleus
$^{40}$Ca, these critical arguments are
basically due to an underestimated role of
the statistical averaging for the level density obtained from the excitation
states and particle number fluctuations.
Another reason is an overestimation
of the shell
model properties of nuclei in a normal state
while we have a wide
superfluid-normal transition beyond the shell model approach.
The MMA 
results were obtained  with only one physical LMSF parameter --
the inverse level density parameter $K$. 
  The 
MMA values
of the inverse level density parameter $K$
for LES's 
can be significantly
 different from those of the neutron resonances.
 Simple estimates of pairing condensation contribution
in spherical magic nuclei at low
excitation energies,
sometimes significantly improve
the comparison with experimental data.

As perspectives, following ideas of Peter Schuck and
    his collaborators,
(see, e.g.,  Ref.~\cite{KR89zpa}), it would be nice to extend  our results
using the super-fluid properties for infinite matter  to those for
finite nuclei. Another fruitful extension, also in line of
activities of Peter Schuck, is applications of our MMA approach
accounting for pairing effects to suitable problems of nuclear astrophysics. 

\vspace{0.2cm}
{\bf Acknowledgments}

\vspace{0.2cm}
 We dedicate this article to the memory of Professor Peter Schuck.

The authors gratefully acknowledge 
A.~Bonasera, M.\ Brack,
D.\ Bucurescu,
A.N.\ Gorbachenko, E.~Koshchiy, S.P.\ Maydanyuk, J.B.~Natowitz, E.~Pollacco,
V.A.\ Plujko, P.\ Ring,
A.\ Volya, and S.\ Yennello
 for many discussions, constructive 
    suggestions, and fruitful help.
This work was partially supported by the
project No. 0120U102221 of
National Academy of Sciences of Ukraine.
S.\ Shlomo and A.G.\ Magner are partially supported by the US
Department of Energy under Grant no. DE-FG03-93ER-40773.

\end{document}